\documentclass[aps,prx,showpacs,twocolumn,amsmath,amssymb,nofootinbib]{revtex4-1}
\usepackage{graphicx}
\usepackage{dcolumn}
\usepackage{bm}
\usepackage{amsmath}
\usepackage{amsfonts}
\usepackage{amssymb}
\usepackage{multirow}

\usepackage{amsthm}
\usepackage{pinlabel}
\usepackage{natbib}

\usepackage{epstopdf}
\usepackage[abs]{overpic}
\usepackage[dvipsnames]{xcolor}

\usepackage{tikz}
\usetikzlibrary{arrows.meta}
\usetikzlibrary{decorations.pathmorphing}
\usetikzlibrary{decorations.markings}

\tikzset{
	particle/.style={thick,draw=blue, postaction={decorate},
		decoration={markings,mark=at position .5 with {\arrow[blue]{triangle 45}}}},
	gluon/.style={decorate, draw=black,
		decoration={coil,aspect=0}}
}

\usepackage{placeins}
\usepackage{cancel}

\usepackage[dvipsnames]{xcolor}

\usepackage{braket}

\definecolor{specialgray}{HTML}{505050}
\definecolor{col10K}{HTML}{FFA000}
\definecolor{col300K}{HTML}{924FA4}
\definecolor{colMu}{HTML}{5278BD}
\definecolor{colMuI}{HTML}{924FA4}

\definecolor{specialgray}{HTML}{505050}
\definecolor{col10K}{HTML}{FFA000}
\definecolor{col300K}{HTML}{924FA4}
\definecolor{colMu}{HTML}{5278BD}
\definecolor{colMuI}{HTML}{924FA4}
\definecolor{newred}{HTML}{D53E4F}
\definecolor{newblue}{HTML}{5278BD}
\definecolor{newcyan}{HTML}{1EA0A0}
\definecolor{newgreen}{HTML}{5CB14E}
\definecolor{newpurple}{HTML}{924FA4}
\definecolor{newyellow}{HTML}{D1C72E}
\definecolor{neworange}{HTML}{D6923C}

\begin{document}

\title{Prediction of an unusual trigonal phase of superconducting LaH$_{\bf 10}$ {stable \\ from 250 to} 425 GPa pressure}
 \author{Ashok K. Verma}\email{hpps@barc.gov.in}
\author{P. Modak}
\affiliation{High Pressure and Synchrotron Radiation Physics Division, Bhabha Atomic Research Centre, Mumbai 400085, India}
\author{Fabian Schrodi}\email{fabian.schrodi@physics.uu.se}
\author{Alex Aperis}
\author{Peter M. Oppeneer}
\email{peter.oppeneer@physics.uu.se}
\affiliation{Department of Physics and Astronomy, Uppsala University, P.\,O.\ Box 516, SE-75120 Uppsala, Sweden}

\vskip 0.4cm
\date{\today}

\begin{abstract}
	\noindent 
	Based on evolutionary crystal structure searches in combination with \textit{ab initio} calculations, we predict an unusual structural phase of  the superconducting  LaH$_{10}$ that is stable from about 250 GPa to 425 GPa pressure. This new phase belongs to a trigonal $R\bar{3}m$ crystal lattice with an atypical cell angle, $\alpha_{rhom}$ $\sim$ 24.56$^{\circ}$. We find that the new structure contains three units of LaH$_{10}$ in its primitive cell, unlike the previously known trigonal phase, where primitive cell contains only one LaH$_{10}$ unit. In this phase, a 32-H atoms cage encapsulates La atoms, analogous to the lower pressure face centred cubic phase. However, the hydrogen cages of the trigonal phase consist of quadrilaterals and hexagons, in contrast to the cubic phase, that exhibits squares and regular hexagons. Surprisingly, the shortest H-H distance in the new phase is shorter than that of the lower pressure cubic phase and of atomic hydrogen metal. We find  a structural phase transition from trigonal to hexagonal at 425 GPa, where the hexagonal crystal lattice coincides with earlier predictions. Solving the anisotropic Migdal-Eliashberg equations we obtain that the predicted trigonal phase (for standard values of the Coulomb pseudopotential) is expected to become superconducting at a critical temperature of about 175 K, which is less than $T_c \sim$250 K measured for cubic LaH$_{10}$. 
\end{abstract}

\maketitle

\section{Introduction}
\label{scIntroduction}
Recent collective efforts in high-pressure experiments and simulations have led to the discovery of a class of superhydride superconductors that to date exhibit the highest critical temperatures ($T_c$) at megabar pressures, see  \cite{Pickard2020,Floreslivas2020} for recent reviews. These superconductors are hydrogen-rich compounds, as for example H$_3$S ($T_c = 203$ K at 150 GPa) \cite{Duan2014,Drozdov2015},  LaH$_{10\pm x}$ ($T_c \approx 250$ K in the pressure region of $137-218$ GPa) \cite{Liu2017,Peng2017,Somayazulu2019,Drozdov2019}, YH$_6$ ($T_c = 203$ K at $166-237$ GPa) \cite{Troyan2020,Kong2019}, ThH$_{10}$ ($T_c = 161$ K at 175 GPa) \cite{Semenok2020}, and the recently discovered carbonaceous sulfur-hydride ($T_c = 287$ K at 267 GPa) \cite{Snider2020}. The underlying mechanism responsible for such high critical temperatures is the conventional electron-phonon coupling as has been discussed recently \cite{Liu2017,Peng2017,Liu2018, Durajski2020, Sun2020, Wang2020, Papaconstantopoulos2020}, even though, as has been noted, the superconducting transition is anomalously sharp \cite{Hirsch2021}.

Lanthanum superhydride has so far provided the highest transition temperature of the rare-earth-hydrides.  Its underlying crystal structure is therefore a topic of concurrent theoretical investigations \cite{Liu2017,Geballe2018,Errea2020,Kruglov2020,Shipley2020}.
Several groups have carried out crystal structure searches, especially in the $100-300$ GPa pressure region. Initial crystal searches  showed that LaH$_{10}$ adopts a stable face centred-cubic lattice, $Fm\bar{3}m$, above 210 GPa, while at lower pressures the cubic phase becomes dynamically unstable \cite{Liu2017}. However, later a combined theoretical and experimental study identified a low symmetry monoclinic structure $C2/m$ as the most likely low pressure phase  \cite{Geballe2018}. In this work it was noticed that, despite the overall monoclinic crystal symmetry, the lanthanum sublattice can be described by a trigonal $R\bar{3}m$ symmetry, which was confirmed by the accompanying x-ray diffraction measurements in decompression experiments \cite{Geballe2018}. Most recently, two additional crystal structures of monoclinic $C2$ and body-centred orthorhombic $Immm$ symmetries 
were added to the list of possible low pressure structures by further crystal structure searches \cite{Errea2020}. In addition, at high pressure ($> 400$ GPa) a hexagonal $P6_3/mmc$ structure was recently predicted \cite{Shipley2020}. It was also noted that the inclusion of anharmonic nuclear quantum corrections reduces the low pressure, low symmetry {$C2$ and $Immm$ structures} to the cubic $Fm\bar{3}m$ structure and thus ruled out the existence of lower symmetry structures for LaH$_{10}$ \cite{Errea2020}. The observed $R\bar{3}m$ lanthanum sublattice, however, yet awaits a satisfactory explanation. In this respect, it is pertinent to mention that former low-symmetry structures are associated with the face-centred cubic lattice through suitable deformations. The crystal structure of LaH$_{10}$ at higher pressures has not yet been fully understood and thus more studies, in this pressure region, offer an exciting possibility for the discovery of new crystal structures.

In this article, we study the structural behavior of LaH$_{10}$ superhydride by performing evolutionary crystal structure searches under pressure, especially, above 250 GPa. We predict a completely new phase that belongs to a trigonal $R\bar{3}m$ crystal symmetry. We analyze the stability of this phase and show that it has a lower enthalpy than the previously predicted face-centred {cubic} $Fm\bar{3}m$ and hexagonal $P6_3/mmc$ phases.
By solving the anisotropic Migdal-Eliashberg equations with \textit{ab initio} input  \cite{Aperis2015} we analyze the superconducting properties and find that the superconducting critical temperature is reduced to $\sim175$ K,  compared to $T_c=250$ K of the cubic phase \cite{Peng2017,Drozdov2019} at lower pressures, for a realistic value of the Coulomb pseudopotential ($\mu^{\star} =0.1$). 
We further find that, although the electron-phonon interaction is responsible for superconductivity, the ratio $\Delta /k_B T_c$ (with $\Delta$ the superconducting gap) deviates from the weak-coupling Bardeen-Cooper-Schrieffer (BCS) value, placing the lanthanum superhydrides in the strong-coupling regime.

\section{Methodology}

To start with, we performed crystal structure searches using the evolutionary algorithm as implemented in the USPEX code \cite{Oganov2006,Lyakhov2013,Oganov2011}. Over the years, this method has been established as a versatile tool for the predictions of  novel stoichiometries and crystal structures of materials at high pressures \cite{Oganov2004,Modak2019,Verma2018,Verma2018_2,Verma2017,Patel2017}. We performed crystal structure searches in the pressure range of $100-500$ GPa using crystal models of one to four formula units of LaH$_{10}$. The first-generation crystal structures are always created randomly, while subsequent generations contain 20\% random structures, and the remaining 80\% of structures are created using heredity, softmutation, and transmutation operators. We computed the crystal structure optimizations, enthalpies and electronic structures within the framework of the density functional theory (DFT) while phonon dispersions and electron-phonon interactions (discussed below) are calculated within the framework of density functional perturbation theory (DFPT). All calculations employed the Perdew-Burke-Ernzerhof version of the exchange-correlation energy functional \cite{Perdew1996}. For  structure optimizations and enthalpies calculations, we use the VASP code \cite{Kresse1994,Bloechl1994,Kresse1996,Kresse1999} and PAW potentials with 600\, eV plane-wave kinetic energy cut-off, and Brillouin zone (BZ) grids of $2\pi\times0.01$ \AA$^{-1}$ interval. 

\begin{figure}[tb!]
	\centering
	\includegraphics[width=0.99\linewidth]{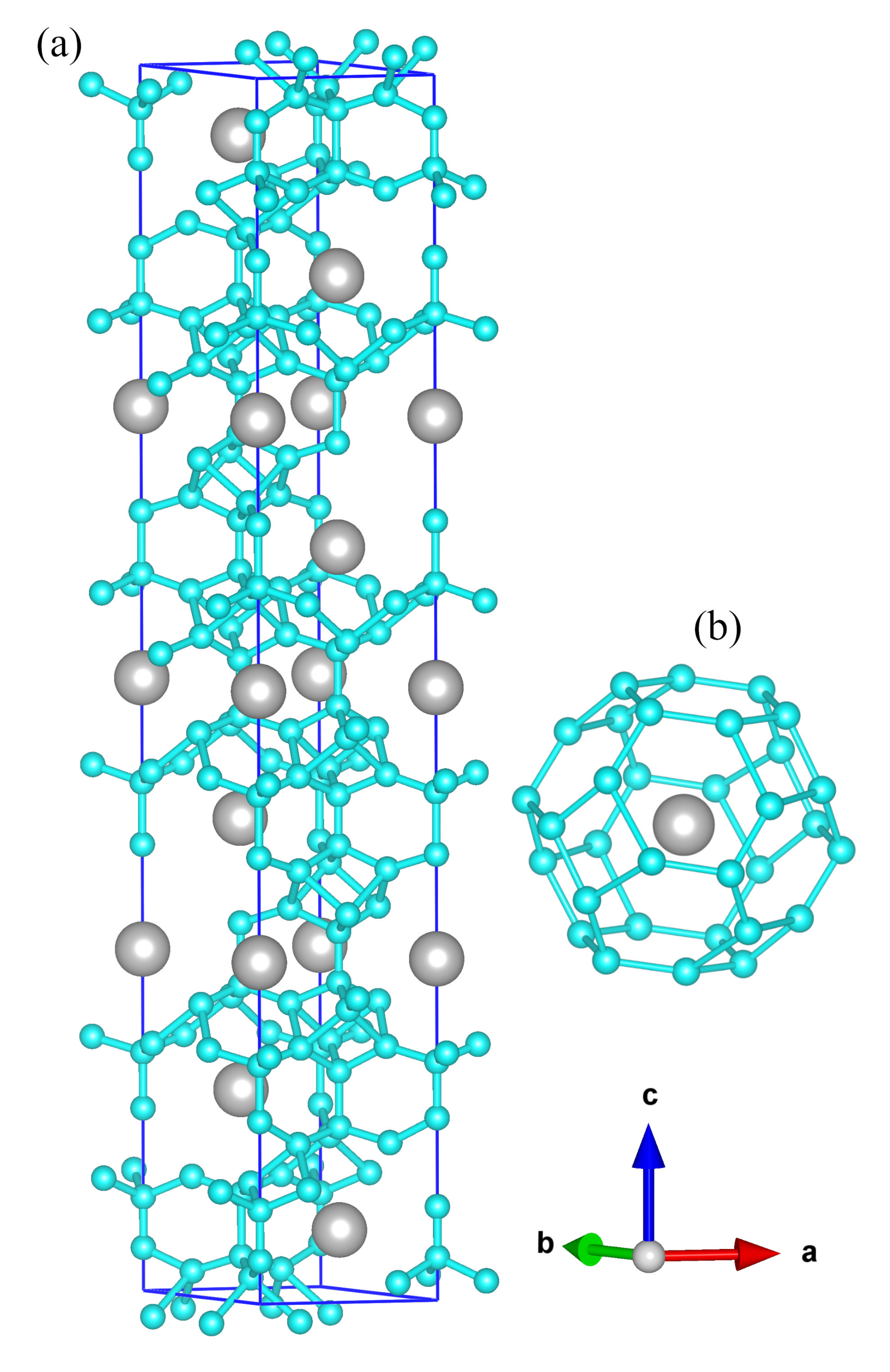}
	\caption {(a) Ball and stick model of the newly discovered trigonal $R\bar{3}m$ phase, in hexagonal setting, of superconducting LaH$_{10}$. (b) A complete model of a 32-H atoms cage around the La atom. Each cage is interlinked to six other cages through cuboids wherein hydrogen atoms occupy the corner positions (not shown here). The 8-hexagonal faces of each 32-H atoms cage are shared by the  surrounding 32-H atoms cages (not shown here). Here, the big grey spheres represent La atoms and small cyan spheres represent H atoms and coordinate system shows orientation of crystal lattice. These structural models are rendered using VESTA software \cite{Momma2011}. } 
\label{structure}
\end{figure}

\section{Results}
\subsection{Structural optimization and stability}

We began with 100 GPa crystal structure searches and these searches readily reproduce previously known structures such as face-centred cubic $Fm\bar{3}m$, trigonal $R\bar{3}m$, monoclinic $C2/m$ and body-centred orthorhombic $Immm$ \cite{Somayazulu2019,Errea2020}. Since the low pressure ($<$ 300\,GPa) phase diagram has been explored extensively in the past by many researchers, we turn our attention to the $300-500$ GPa pressure region.  
Our crystal structure searches in this region produce two crystal structures, namely a trigonal structure $R\bar{3}m$ and a hexagonal structure $P6_3/mmc$. The primitive cell of our trigonal structure, {shown in Fig.\ \ref{structure})} consists of \textit{three} formula units of LaH$_{10}$, unlike  the previously known trigonal structure which consists of only one formula unit of LaH$_{10}$ \cite{Liu2017,Geballe2018,Errea2020}. We also notice that the new trigonal structure has an anomalously small cell angle, $\alpha_{rhom}$ $\sim$ 24.56$^{\circ}$, unlike the earlier trigonal structure for which $\alpha_{rhom}$ $\sim$ 60$^{\circ}$ \cite{Liu2017,Errea2020}.
In this trigonal phase, the lanthanum atoms are surrounded by cages consisting of 32-H atoms, each of which is linked to six neighboring cages via cuboids of 8-H atoms. The 8-hexagonal faces of each 32-H atoms cage are shared by the surrounding 32-H atoms cages. Contrary to the low pressure cubic phase, the hydrogen cage in the trigonal structure is made of quadrilaterals and hexagons, as shown in Fig.\ \ref{structure}.
 
\begin{figure}[tb!]
	\centering
	\includegraphics[width=1.08\linewidth]{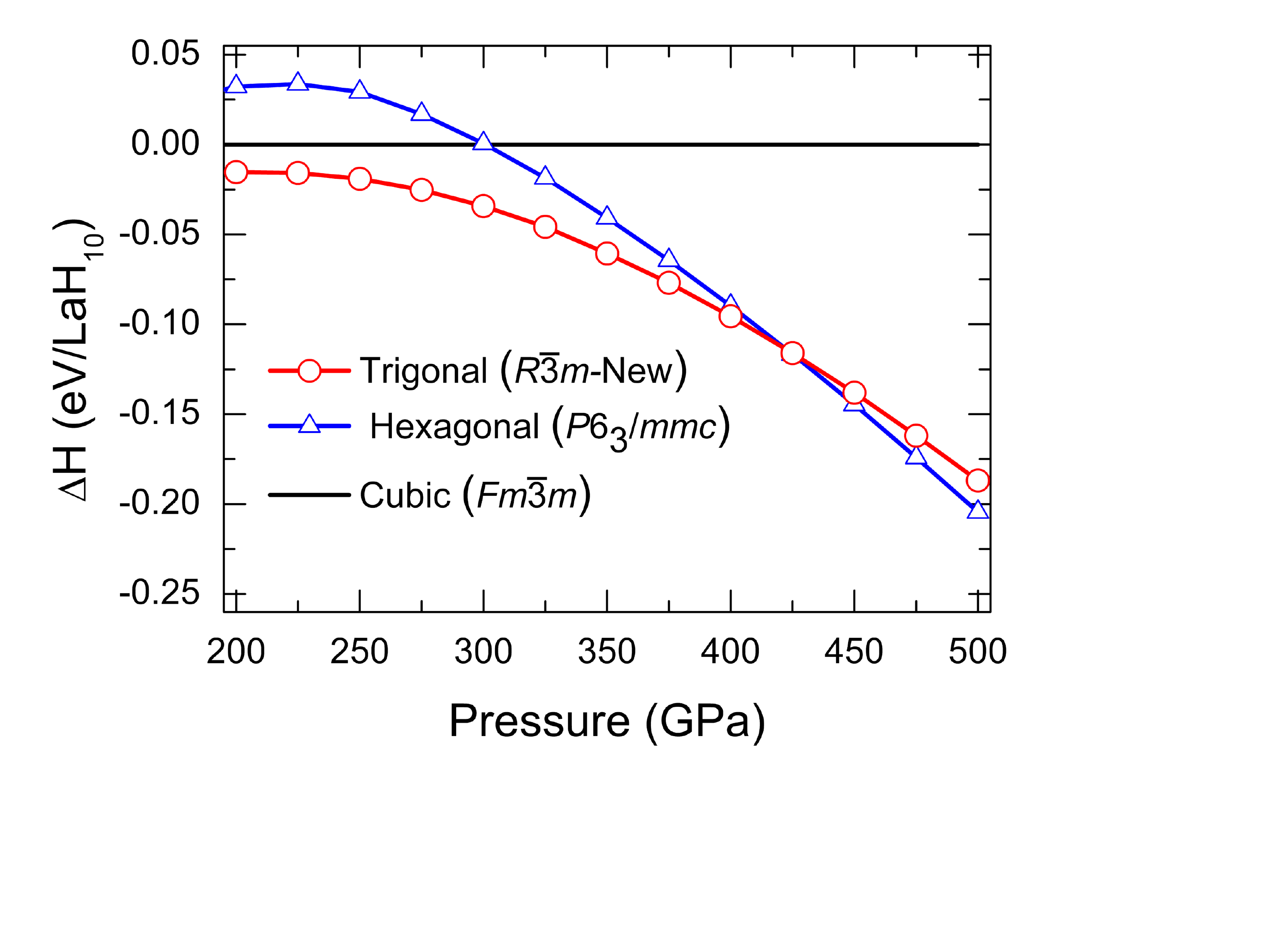}
		\vspace*{-1.7cm}
	\caption{Computed enthalpy difference of trigonal and hexagonal structures of LaH$_{10}$ with reference to the face-centred cubic structure.}
	\label{enthalpy}
\end{figure}

Interestingly,  the new trigonal phase has slightly lower enthalpy, $< 2.0$ meV$/$atom, than  the face-centred cubic phase even  for lower pressures, $<$ 250 GPa, thus  making these phases  energetically indistinguishable in this pressure region. In addition, the earlier mentioned argument, that nuclear quantum corrections  (zero-point vibrations) of hydrogen destabilize low-symmetry structures in favor of the cubic $Fm\bar{3}m$ structure \cite{Errea2020} can play a role here. However, the enthalpy difference between trigonal and cubic phases grows with increasing pressure, reaching to a value of $\sim$ 9.0 meV$/$atom at 400 GPa, as shown in Fig.\ \ref{enthalpy}. We also find that the new trigonal phase transforms to a hexagonal phase above 425 GPa. Our hexagonal phase belongs to the same crystal lattice ($P6_3/mmc$) as that of previous works \cite{Shipley2020}. In this study, we have not attempted to estimate the phonon contributions of the free energies due to the involvement of computationally expensive phonon calculations at several pressures for all candidate structures.
For the trigonal phase, we obtain  (in hexagonal setting) $a = b = 3.812$ {\AA} and $c = 23.0467$ {\AA} at 350 GPa. Corresponding Wyckoff positions are, La1: $3b$: (0.0000, 0.0000, 0.5000), La2: $6c$: (0.0000, 0.0000, 0.7213), H1: $6c$: (0.0000, 0.0000, 0.8055), H2: $6c$: (0.0000, 0.0000, 0.4147), H3: $18h$: (08276, 0.1734, 0.2059),  H4: $18h$: (0.1656, 0.8343, 0.0944), H5: $18h$: (0.1658, 0.8342, 0.0174), H6: $6c$: (0.0000, 0.0000, 0.0348), H7: $6c$: (0.0000, 0.0000, 0.6356), H8: $6c$: (0.0000, 0.0000, 0.8513), H9: $6c$: (0.0000, 0.0000, 0.0766). 

\begin{figure}[tb!]
	\centering
	\includegraphics[width=1\linewidth]{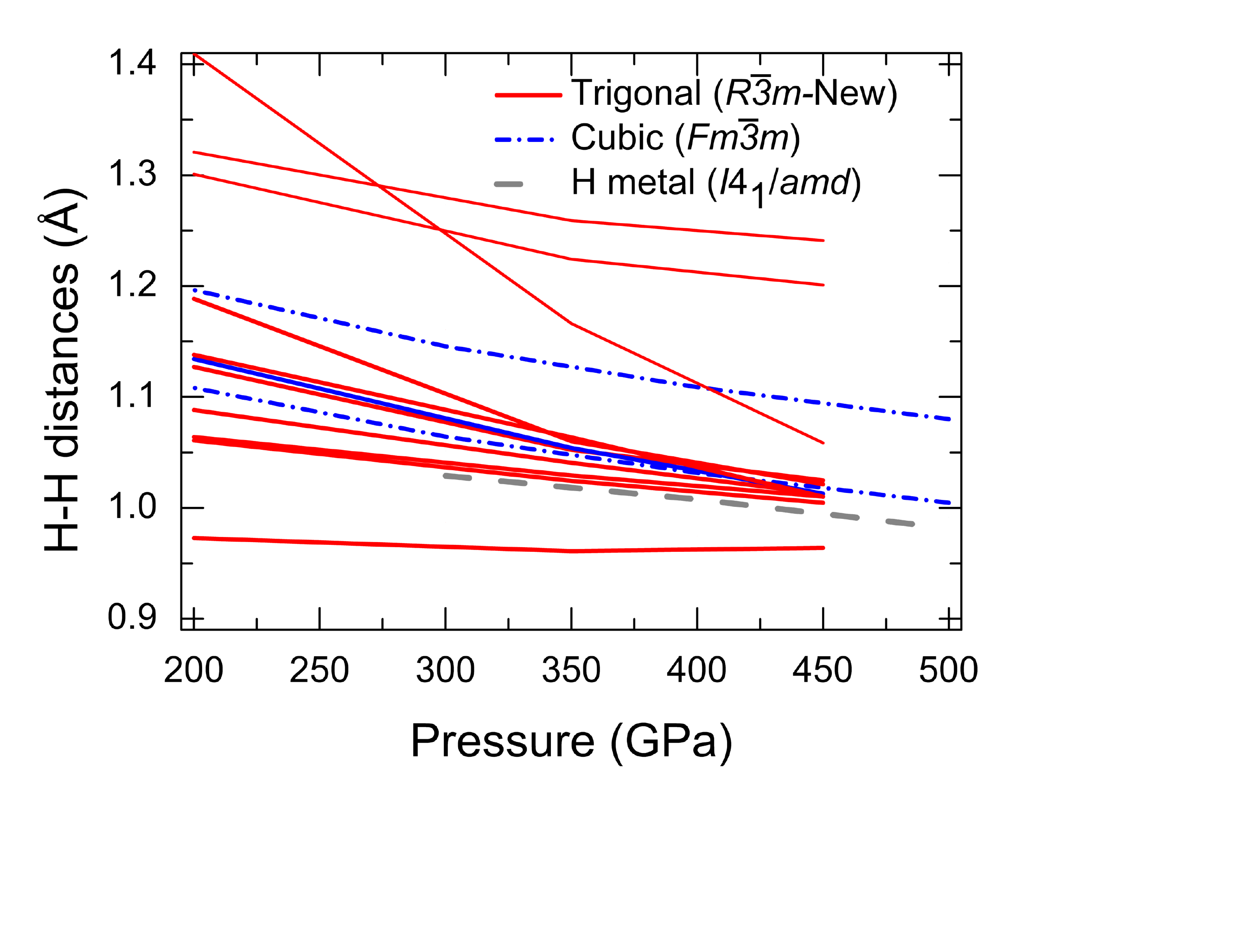}
		\vspace*{-1.7cm}
	\caption{Comparison of H-H distances as function of pressure in the face-centred cubic and trigonal structures along with atomic hydrogen metal \cite{verma2020}.}
	\label{H-H}
\end{figure}

We did not find any other lower enthalpy structure up to pressures of 500 GPa. Similar to the cubic $Fm\bar{3}m$ phase the trigonal phase also has a 32-atoms hydrogen cage around the La atoms, see  Fig.\ \ref{structure}(b). However, square and hexagonal faces are now distorted, probably to accommodate the symmetry changes. Here, the square and regular hexagonal faces of the cubic H-cage deform into quadrilateral and irregular hexagonal faces \cite{Somayazulu2019}. As shown in Fig.\ \ref{H-H}, this leads to a splitting of the two H-H distances of the face-centred cubic phase into many different H-H distances. Notably, the smallest H-H distances are smaller than those of the cubic phase as well as those of hydrogen metal at similar pressures \cite{verma2020}. 

\begin{figure}[tb!]
	\centering
	\includegraphics[width=1\linewidth]{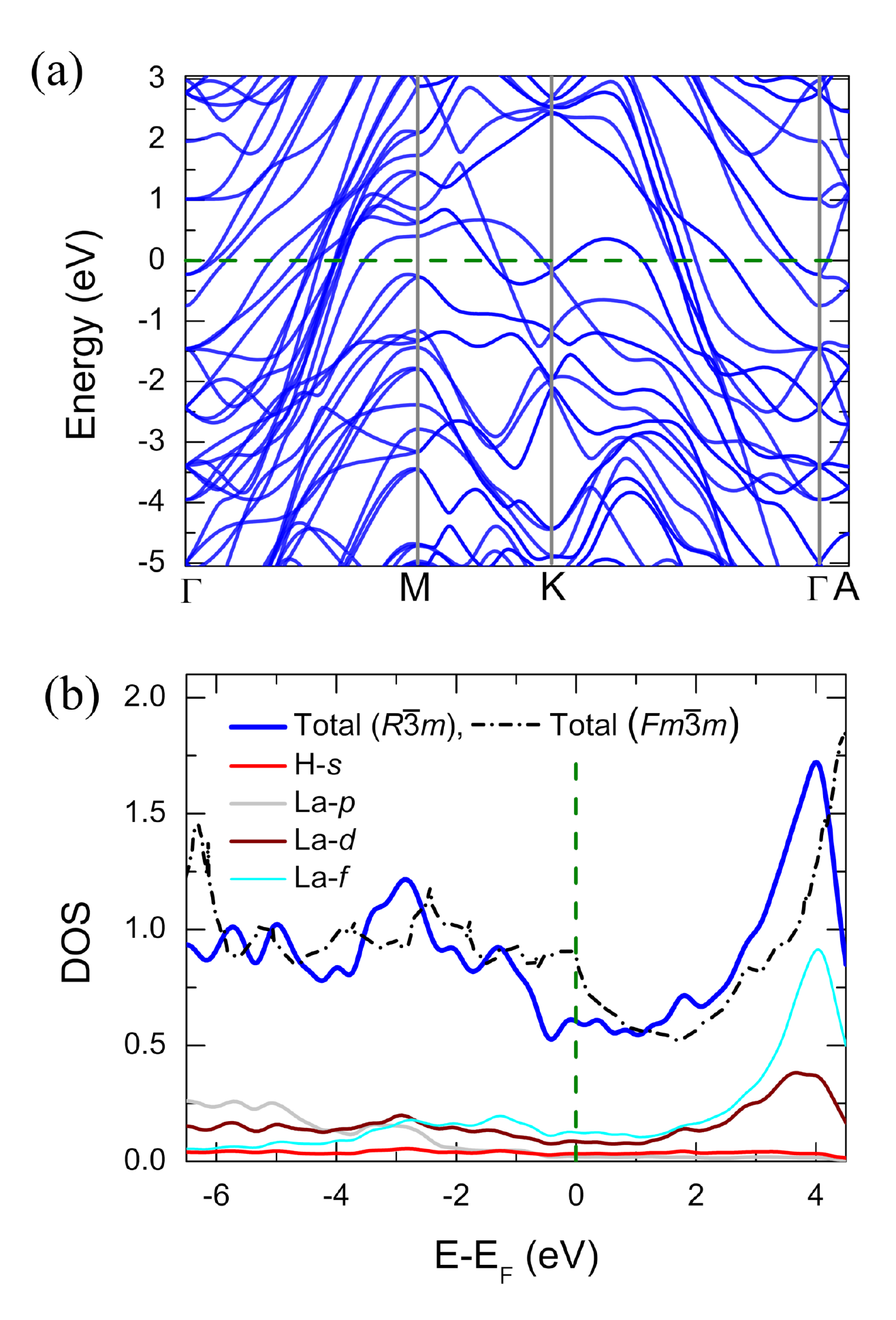}
	\vspace*{-0.7cm}
	\caption{(a) Electronic band structure of the trigonal $R\bar{3}m$ structure, in hexagonal setting, along the high-symmetry directions of the Brillouin zone at 350 GPa. (b) Total density of states (DOS) and atom projected partial density of states (PDOS). Here, total DOS is given in units of states/eV per LaH$_{10}$ unit and PDOSs are given in units of states/eV/atom. Contributions of La-$s$ states are insignificant to the total DOS values at Fermi level, and hence not shown. The La and  H  PDOSs represent the average contribution of 2- and 9-types of La and H atoms, respectively. For comparison, the total DOS for the cubic $Fm\bar{3}m$ phase is also shown. Vertical and horizontal broken lines in panels (a) and (b), respectively, show the position of the Fermi level.}
	\label{band}
\end{figure}

We now turn our attention to the electronic and phonon properties of the new phase. These were calculated using DFPT as implemented in the Quantum Espresso package \cite{Giannozzi2009}. We use ultrasoft pseudopotentials with 50 and 500 Ry cut-off for plane-wave kinetic energy and charge density, respectively. We used a $24 \times 24 \times 24$ Monkhorst-Pack \cite{Monkhorst1976} $\mathbf{k}$-point grid for electronic properties and  a $12\times12\times12$
$\mathbf{k}$-point mesh for phonons. Force constants, phonons, and electron-phonon couplings were calculated on a $4\times 4 \times 4$ $\mathbf{q}$-point mesh. A test run with denser $36 \times 36 \times 36$ $\mathbf{k}$-point meshes does not show significant changes.

In Figure \ref{band} we show the electronic band structure and density of states (DOS) at 350 GPa. The electronic band structure clearly shows that the trigonal phase is a good metal, similar to the lower pressure cubic phase. Many electronic bands cross the Fermi level along various directions of the BZ, see Fig.\ \ref{band}(a). The orbital projection of the DOS  reveals that La-$d$ and La-$p$ states contribute most to the DOS at the Fermi level, while La-$s$ contributions are insignificant and hence not shown in the plot. Similar DOS trends were also noticed for the cubic phase \cite{Liu2017}. However, the trigonal phase has considerably smaller DOS values ($N_0$) at the Fermi level than the cubic phase. For example, at 350 GPa we compute $N_0=0.60$ states/eV and $N_0=0.88$ states/eV {per LaH$_{10}$ unit} for the trigonal and cubic phases, respectively.  

\begin{figure}[tb!]
	\flushleft
	\hspace*{-0.3cm}
	\includegraphics[width=1.05\linewidth]{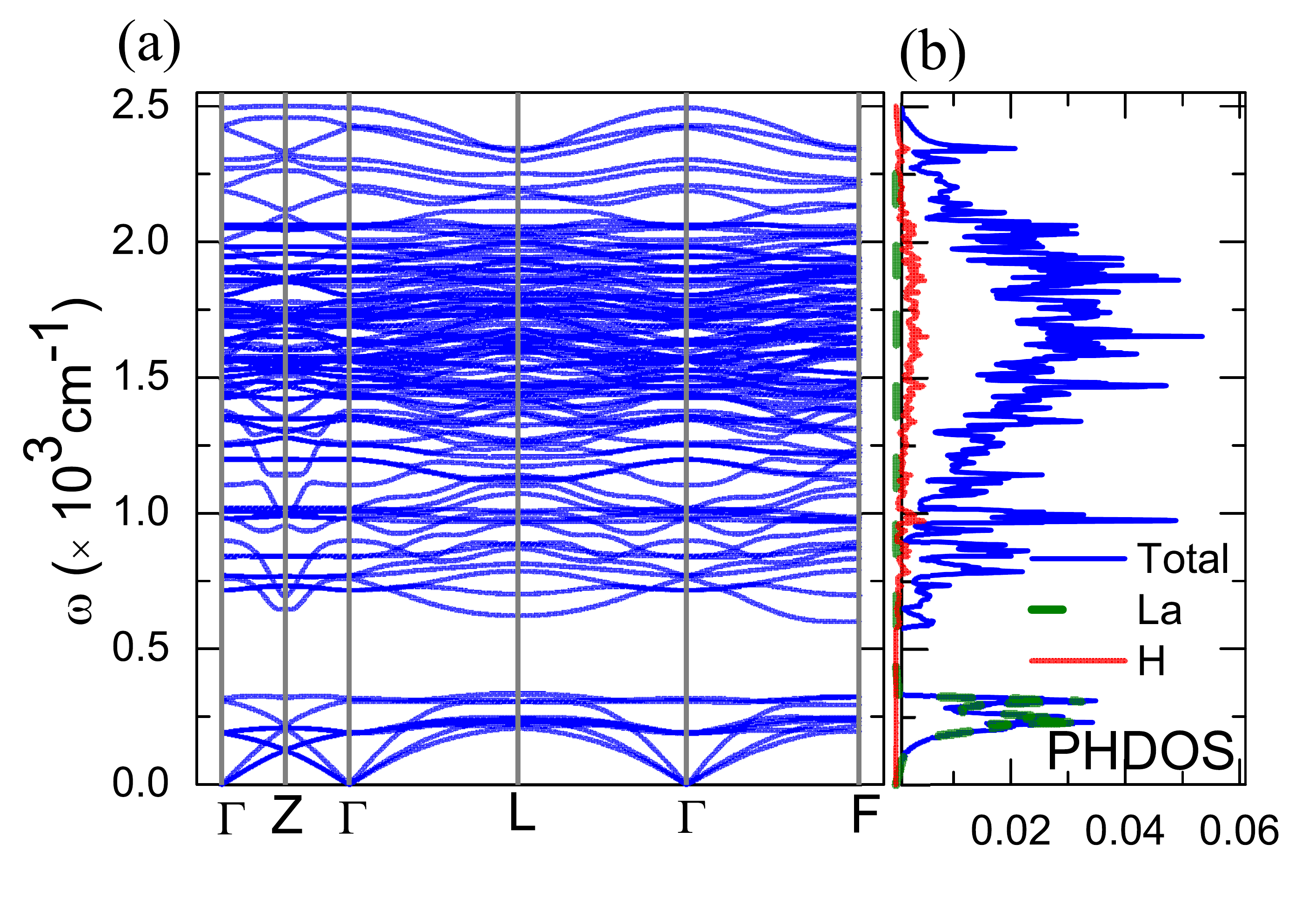}
	\vspace*{-0.7cm}
	\caption{(a) Phonon dispersions computed for the primitive trigonal unit cell along high-symmetry lines of the Brillouin zone. (b) The corresponding phonon density of states (PHDOS).  Here, the total phonon DOS, represents contribution of a single LaH$_{10}$ unit. The La and H  PHDOS represent values per atom which are computed by performing averages over 2- and 9-types of La and H atoms, respectively.} 
	\label{Phonon}
\end{figure}

The phonon dispersions of the trigonal phase are shown in Fig.\,\ref{Phonon}(a) along high symmetry lines of the BZ, and the respective phonon density of states (PHDOS) in panel (b) of the same graph. It is pertinent to mention that the highest phonon frequency, $\approx 2500$ cm$^{-1}$, of this phase lies between the highest frequencies  of  the cubic phase and of atomic hydrogen metal. These phonon frequencies are  $\approx 2000$  cm$^{-1}$ at 300 GPa \cite{Liu2017} and  $\approx 2600$  cm$^{-1}$ at 400 GPa \cite{verma2020} for the cubic LaH$_{10}$ and hydrogen metal, respectively. 
Evidently, the La atoms contribute mainly to the low frequency part of the phonon spectrum, below 500 cm$^{-1}$, whereas H atoms contribute mainly to the high frequency part of the phonon spectrum, as can be recognized in Fig.\ \ref{Phonon}(b).

\subsection{Superconducting properties}

Next, we calculate the superconducting properties of the here-discovered trigonal phase at 350 GPa by numerically solving the anisotropic Migdal-Eliashberg equations,
\begin{align}
\!\!\!\!Z_{\mathbf{k},m} \!=&  1 + \frac{\pi T}{\omega_m} \! \sum_{\mathbf{k}',m'} \!\frac{\delta(\xi_{\mathbf{k}'})}{N_0} \lambda_{\mathbf{k}-\mathbf{k}',m-m'} \frac{\omega_{m'}}{\sqrt{\omega_{m'}^2 +\Delta_{\mathbf{k}',m'}^2}}, \label{zz}
\\
\Delta_{\mathbf{k},m} =& \frac{\pi T}{Z_{\mathbf{k},m}} \sum_{\mathbf{k}',m'} \frac{\delta(\xi_{\mathbf{k}'})}{N_0} [ \lambda_{\mathbf{k}-\mathbf{k}',m-m'}- \mu^{\star}(\omega_c)] \nonumber\\
& ~~\times\frac{\Delta_{\mathbf{k}',m'}}{\sqrt{\omega_{m'}^2 +\Delta_{\mathbf{k}',m'}^2}} ,\label{delta}
\end{align}
as implemented in the Uppsala Superconductivity  code (\textsc{uppsc}) \cite{UppSC,Aperis2015,Schrodi2019,Schrodi2020_2,Schrodi2020_3}. In the above, $Z_{\mathbf{k},m}$ and $\Delta_{\mathbf{k},m}$ are the mass renormalization and superconducting gap function, respectively.  Equations (\ref{zz}) and (\ref{delta}) are solved self-consistently in Matsubara space (with fermionic Matsubara frequency $\omega_m=\pi T(2m+1)$, $m \in\mathbb{Z}$), as function of temperature $T$ and Anderson-Morel Coulomb pseudopotential $\mu^{\star}$. The dynamic electron-phonon coupling is calculated via $\lambda_{\mathbf{q},l}=\sum_{\nu}\lambda_{\mathbf{q},\nu}\omega_{\mathbf{q},\nu}^2/(\omega_{\mathbf{q},\nu}^2+q_l^2)$, {with $q_l=2\pi Tl$ ($l \in\mathbb{Z}$) the bosonic Matsubara frequencies.} The electron energies $\xi_{\mathbf{k}}$, density of states at the Fermi level $N_0$, phonon frequencies $\omega_{\mathbf{q},\nu}$ ($\nu$ branch index) and electron-phonon couplings $\lambda_{\mathbf{q},\nu}$ are obtained from the above-described {\em ab initio} calculations. The total electron-phonon coupling $\lambda$ is given by $\lambda = \sum_{\mathbf{q},\nu} \lambda_{\mathbf{q},\nu}$.

\begin{figure}[t!]
	\centering
	\includegraphics[width=1\linewidth]{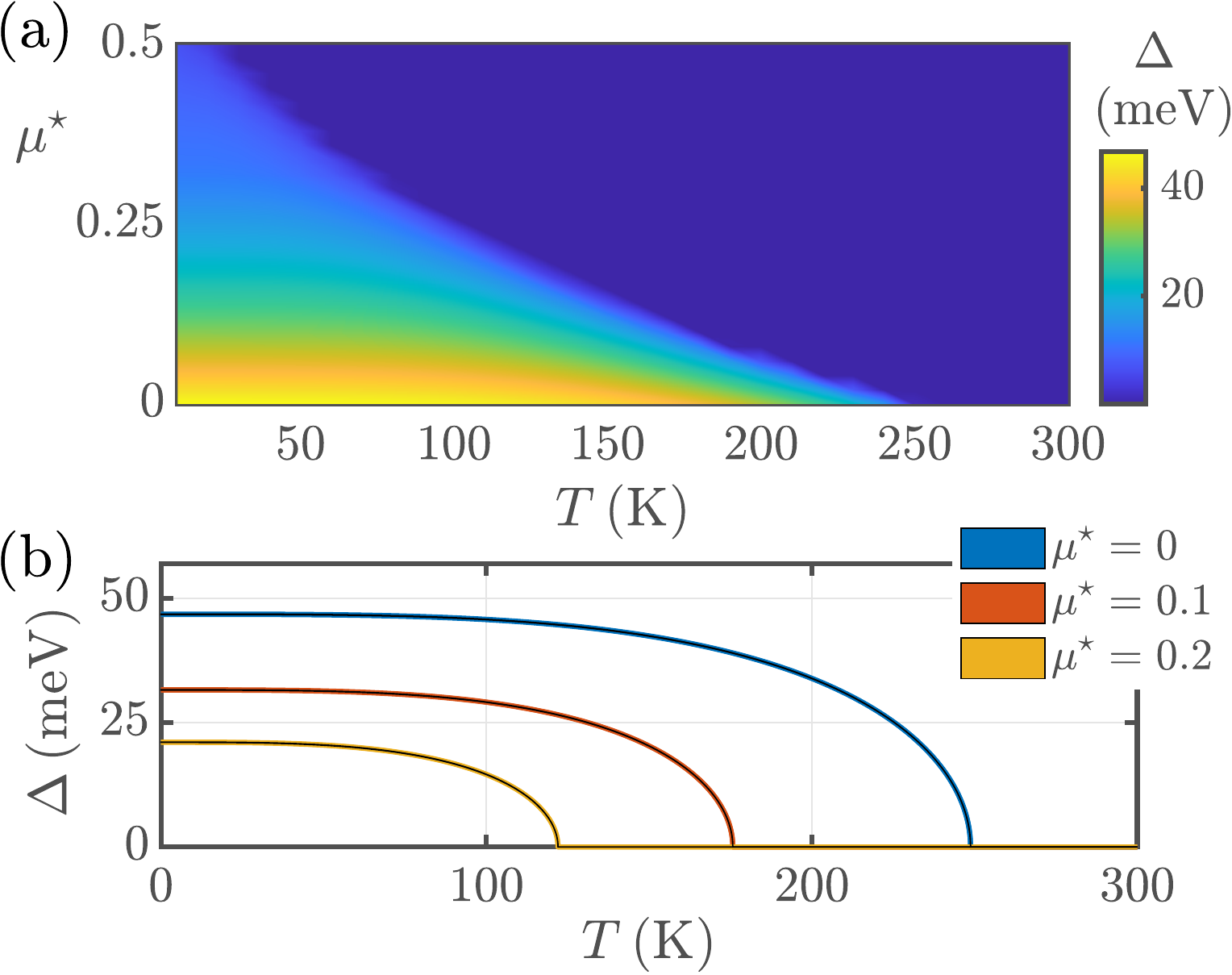}
	\caption{(a) Selfconsistently calculated maximum superconducting gap $\Delta$ as a function of temperature and screened Coulomb potential $\mu^{\star}$. (b) Temperature dependent $\Delta$ computed for three values of $\mu^{\star}$ as written in the legend. The thickness of each curve reflects the momentum anisotropy of the superconducting gap.}\label{gap}
\end{figure}

In Fig.\,\ref{gap}(a) we show the maximum zero-frequency superconducting gap $\Delta=\underset{\mathbf{k}}{\mathrm{max}}\,\Delta_{\mathbf{k},m=0}$ as function of $T$ and $\mu^{\star}$. For small values of the Coulomb pseudopotential, $\Delta$ reaches around {45} meV in the limit $T\rightarrow0$. Considering a very wide range of $\mu^{\star}$, the critical temperatures range from 30 K to approximately 250 K.
For closer inspection we choose three specific values $\mu^{\star}\in\{0,0.1,0.2\}$ that are close to the archetypal value $\mu^{\star}=0.1$ and plot the temperature dependence of $\Delta$ in Fig.\ \ref{gap}(b). As a measure of momentum  anisotropy of the gap, we show here the range $[\underset{\mathbf{k}}{\mathrm{min}}\,\Delta_{\mathbf{k},m=0}, \underset{\mathbf{k}}{\mathrm{max}}\,\Delta_{\mathbf{k},m=0}]$ for each $\mu^{\star}$ and $T$. As is directly apparent, each shaded area in Fig.\ \ref{gap}(b) is very thin, reflecting a negligible degree of gap anisotropy, a property which goes in line with results for most hydride superconductors. The critical temperature for the typical value of $\mu^{\star}=0.1$ \cite{Liu2017,Peng2017,Liu2018,Sun2020} is approximately {175} K, which is a significant decrease compared to $T_c\leq250$ K in the lower-pressure cubic phase of LaH$_{10}$ \cite{Liu2017,Drozdov2019}. This change in $T_c$ can partially be understood in terms of the different values for {the DOS at the Fermi energy,} $N_0$, in the 
trigonal and cubic phase, as was mentioned before. Additionally, the electron-phonon coupling strength $\lambda=1.69$ as found here is smaller than the $\lambda=2.2$ that has been estimated for the cubic phase at 250 GPa \cite{Liu2017}. It deserves further to be mentioned that suitably doping of LaH$_{10}$ could provide an increase of the Fermi-energy DOS and hence enhance $T_c$ \cite{Floreslivas2020b}.

In Table \ref{table} we list some characteristics of the superconducting state for the same values of the Coulomb pseudopotential. The limiting values $\Delta(T=0)$ have been obtained via the fitting function $\Delta(T)\simeq{\rm Re}\sqrt{\alpha - T^{\beta}/\gamma}$. Remarkably, for all choices of $\mu^{\star}$ we find the ratio $\Delta/k_BT_c$ in the strong-coupling regime, in contrast to the weak-coupling BCS value of 1.76. Thus, although the electron-phonon interaction is prevalent, lanthanum superhydride cannot be considered as a BCS superconductor.\\

\renewcommand{\arraystretch}{1.4}
\begin{table}[t!]
\begin{ruledtabular}
	\caption{Characteristic superconducting properties of trigonal LaH$_{10}$ computed for different values of the Coulomb pseudopotential $\mu^{\star}$ at a pressure of 350 GPa.}\label{table}
	\begin{tabular}{cccc } 
		~~$\mu^{\star}$~~ & ~~$T_c ~(\mathrm{K})$~~ & ~~$\Delta(T=0)~(\mathrm{meV})$~~ & ~~$\Delta/k_BT_c$~~  \\ 
		\hline
		$0$   & {$248.7$} & {$46.9$} & {$2.19$}   \\
		$0.1$ & {$175.5$} & {$31.6$} & {$2.09$}   \\
		$0.2$ & {$122.0$} & {$21.1$} & {$2.01$}   \\
	\end{tabular}
	\end{ruledtabular}
\end{table} 

\FloatBarrier

\section{Conclusions}

In summary, we have predicted a new trigonal phase of superconducting LaH$_{10}$ at pressures above 250 GPa. The new  phase has  an anomalously small cell angle and its primitive cell is made of three units of LaH$_{10}$, unlike the previously known trigonal phase where only one unit of LaH$_{10}$ resides in the primitive cell. Analogous to the lower pressure cubic phase, this phase also has a 32-H atoms cage encapsulating the lanthanum atom and each cage is interconnected to the neighboring six such cages. However, the crystal symmetry reduction from cubic to trigonal 
results in significant distortions of the hydrogen cages around the La atoms. Interestingly, the smallest H-H distances in the trigonal phase are smaller than those of the cubic phase and those of atomic hydrogen metal. We have also found that the trigonal phase transforms to a hexagonal phase, where the crystal lattice is identical with that given in earlier predictions. Solving the anisotropic Migdal-Eliashberg equations on the basis of \textit{ab initio} input we predict strong-coupling superconductivity with a high transition temperature, $T_c \approx {175}$ K for the archetypal value $\mu^{\star}=0.1$, at 350 GPa, a $T_c$ value which is thus somewhat lower than $T_c \approx 250$ K that was measured \cite{Drozdov2019} and computed \cite{Liu2017} for the cubic LaH$_{10}$ phase.

\begin{acknowledgments}
A.K.V.\ and P.M.\ acknowledge the support of ANUPAM supercomputing facility of BARC. 
F.S., A.A., and P.M.O.\ acknowledge support by the Swedish Research Council (VR), the R{\"o}ntgen-{\AA}ngstr{\"o}m Cluster, and the Knut and Alice Wallenberg Foundation (No.\ 2015.0060). {The Eliashberg-theory calculations were enabled by resources provided by the Swedish National Infrastructure for Computing (SNIC) at NSC Link\"oping, partially funded by VR through Grant Agreement No.\ 2018-05973.}
\end{acknowledgments}

\bibliographystyle{apsrev4-1}


%

\end{document}